\documentclass[a4paper,11pt]{article}

\usepackage[utf8]{inputenc}
\usepackage[T1]{fontenc}
\usepackage{lmodern}
\usepackage{geometry}
\usepackage{graphicx}
\usepackage{authblk}
\usepackage{url}
\usepackage{cite}
\usepackage{fancyhdr}
\usepackage[hidelinks]{hyperref}
\usepackage{orcidlink}

\newcommand{\tauh}{$\tau_\mathrm{h}$ }
\newcommand{\pt}{$p_{\mathrm{T}}$ }
\newcommand{\tauhns}{$\tau_\mathrm{h}$}
\newcommand{\ptns}{$p_{\mathrm{T}}$}
\newcommand{\fbinv}{$\,\mathrm{fb}^{-1}$}

\newenvironment{tight_itemize}{
	\begin{itemize}
		\setlength{\itemsep}{0.5pt}
		\setlength{\parskip}{0.3pt}
	}{\end{itemize}}

\fancypagestyle{firstpage}{
	\fancyhf{}
	\fancyhead[C]{\includegraphics[height=2.2cm]{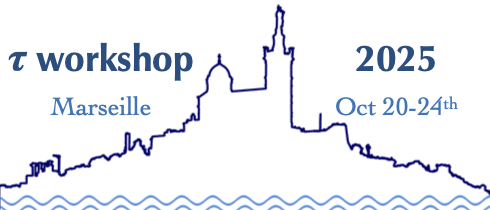}\\ \textit{Proceedings of the 18th International Workshop on Tau Lepton Physics}
	} 

}

\title{Recent advancements in the tau reconstruction and identification techniques in CMS}
\author[1]{Andrea Cardini\,\orcidlink{0000-0003-1803-0999}\thanks{Presenting author.}} 

\affil[1]{ICTEA--Universidad de Oviedo, Asturias, Spain}

\date{}

\begin{document}

\maketitle
\thispagestyle{firstpage}

\vspace{-1cm}
\begin{center}
	\textit{On behalf of the CMS Collaboration}
\end{center}
\vspace{0.5cm}

\begin{abstract}
	Tau leptons play a crucial role in studies of the Higgs boson and searches for Beyond the Standard Model physics at the present LHC and in its high luminosity upgrade. This talk presents the latest advancements in the reconstruction and identification of hadronic decays of tau leptons at the CMS experiment, both at the online and offline levels. The tau identification algorithm deployed for the early Run 3 data-taking period, based on a deep convolutional neural network with domain adaptation, showcases significantly improved discrimination of genuine hadronic tau decays against mis-identified quark and gluon jets, electrons, and muons. During live data-taking, a simplified version of the algorithm is used to select events with tau leptons at the High Level Trigger (HLT). The performance and calibration of both algorithms using early Run 3 data are presented. Many CMS physics analyses involving tau leptons are expected to benefit from these improvements. Alternative approaches to identify hadronic taus combined with jet flavour, based on graph neural networks and particle transformers, are also covered. Additionally, the dedicated techniques used to reconstruct and identify displaced tau leptons originating from long-lived particle decays using graph neural networks are discussed.
\end{abstract}

\section{Introduction}

The CMS Collaboration~\cite{CMS:2008xjf,CMS:2023gfb} at the CERN LHC has a long history of physics analyses requiring the precise and accurate reconstruction of tau leptons. These include measurements of the Higgs boson production~\cite{HIG-16-043,HIG-19-010} and properties under CP symmetry~\cite{HIG-20-006}, which relies on the identification of the tau decay products and their angular distributions~\cite{Cardini:2022qqy}, the measurement of the tau polarization~\cite{SMP-18-010} and anomalous magnetic moment~\cite{HIN-21-009,CMS-PAS-HIN-24-011}, and searches for new physics in final states with tau leptons~\cite{HIG-21-001,PAS-EXO-24-012,PAS-EXO-24-020}. 
This rich program is supported by continuous advancements in reconstruction and identification techniques deployed by the CMS Collaboration, the most recent of which are presented in these proceedings. 

\vspace{-0.3cm}
\section{Tau Reconstruction and Identification in CMS}
\label{sec:2overview}

Tau leptons produced in ``standard'' topologies, i.e., prompt production via Drell--Yan, Yukawa, or charge current electroweak interactions, generally decay before reaching the innermost layers of the CMS detector~\cite{CMS:2008xjf,CMS:2023gfb}. They are therefore reconstructed from their decay products.
Leptonically decaying tau leptons are reconstructed following the standard reconstruction techniques for prompt electrons and muons~\cite{CMS:2017yfk,CMS:2020uim,CMS:2018rym,CMS:2014pgm}; their identification is inferred based on the global event topology (properties of the dilepton system, presence of energy imbalance in the transverse plane, etc.). 
Isolated hadronically decaying tau leptons (\tauhns) candidates are reconstructed by the hadron-plus-strip (HPS) algorithm~\cite{CMS:2018jrd}, which selects a combination of 1 or 3 tracks with energy deposits in the calorimeters, to identify the tau decay channel. Neutral pions are reconstructed as strips with dynamic size in $\eta$-$\phi$ from the reconstructed electrons and photons. The HPS algorithm was tuned to prioritize the efficiency of the reconstruction and is used in conjunction with dedicated algorithms to identify genuine \tauh decays from jets originating from the hadronization of quarks or gluons, and from electrons or muons.
The convolutional neural network (CNN) \textsc{DeepTau} has been used~\cite{CMS:2022prd} for this purpose since the end of the LHC Run 2. Information from all individual reconstructed particles near the \tauh axis is combined with properties of the \tauh candidate and the event.

\subsection{Recent Advancements in Tau Identification}
\begin{figure}[h]
    \centering
    \includegraphics[width=0.7\linewidth]{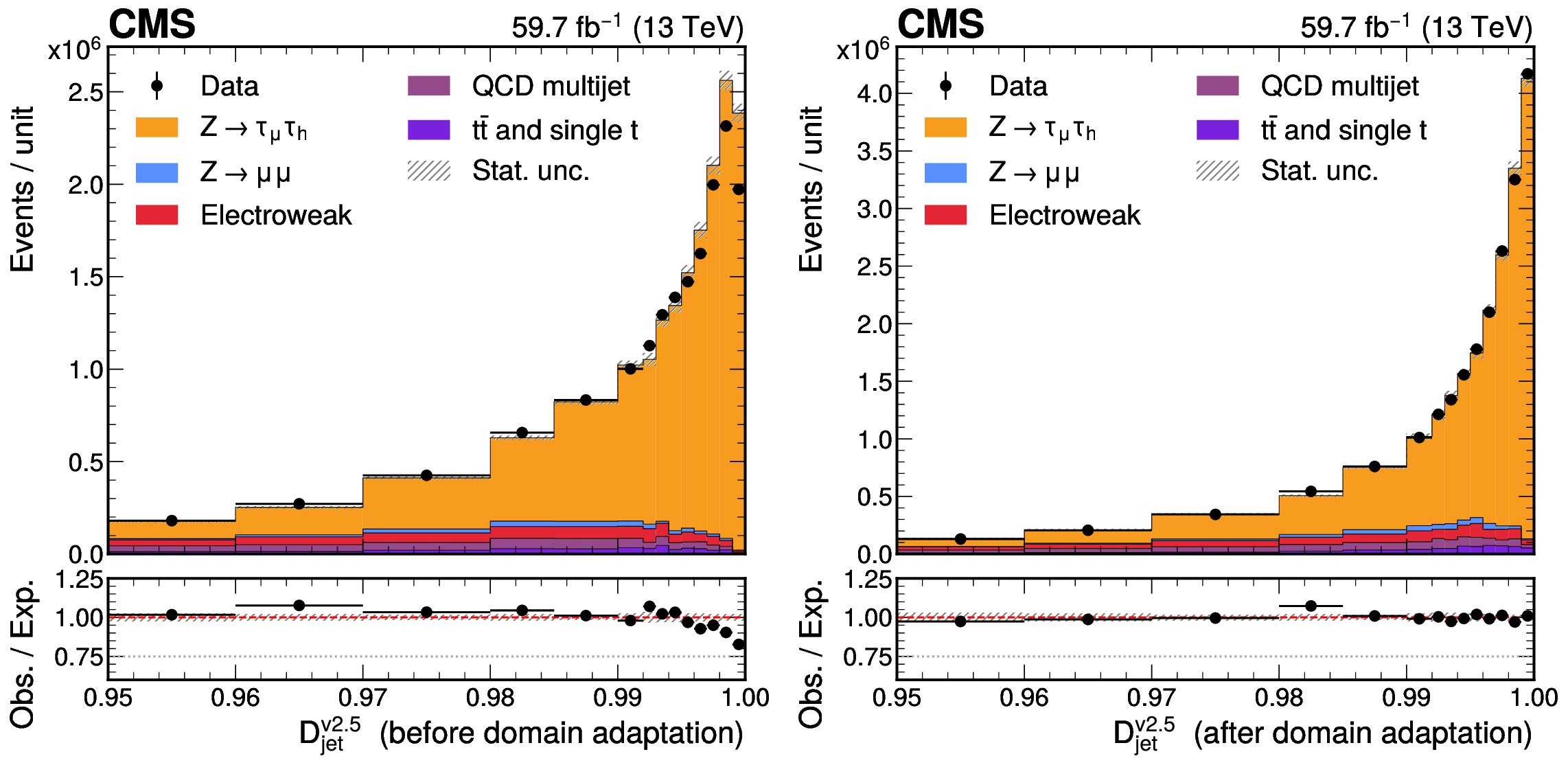}
    \caption{Distribution of the \textsc{DeepTau} discriminator against quark and gluon jets before (left) and after (right) domain adaptation, for the dataset used to train the domain adaptation~\cite{CMS:2025kgf}.}
    \label{fig:domainadapt}
\end{figure}

\begin{figure}[h]
    \centering
    \includegraphics[width=0.35\linewidth]{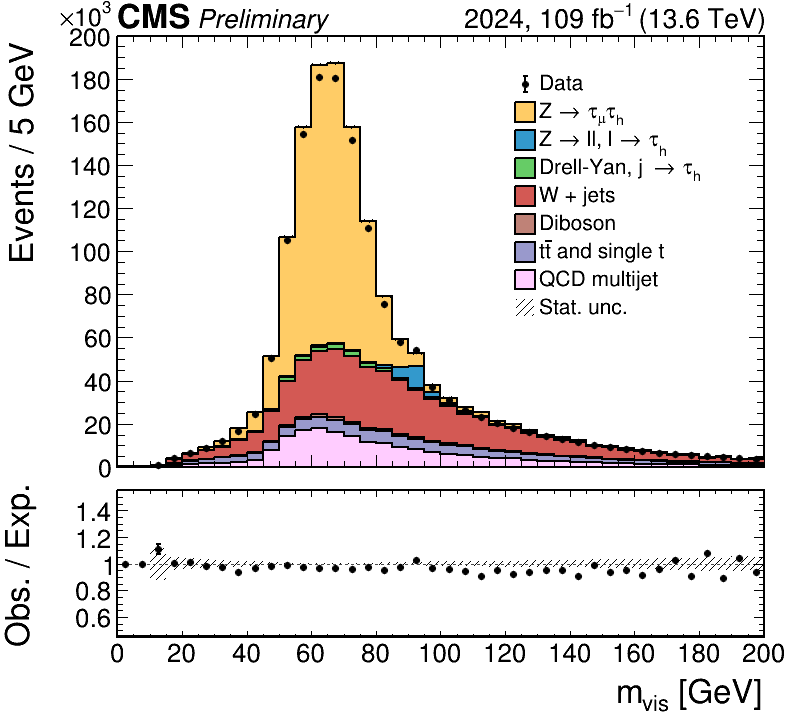}
    \includegraphics[width=0.35\linewidth]{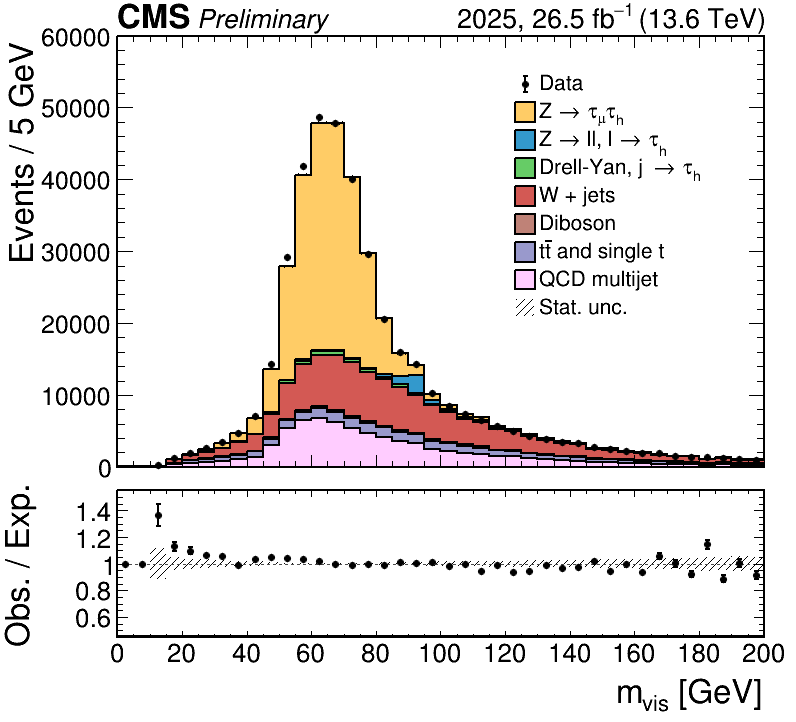}
    \caption{Visible mass distribution for a muon and \tauh system for 2024 (left) and 2025 (right) data~\cite{CMS-DP-2025-074}. For 2024, data correspond to an integrated luminosity of 109\fbinv. For 2025, data correspond to an integrated luminosity of 26.5\fbinv, collected up to the 23$^\mathrm{rd}$ of July 2025.}
    \label{fig:2024-5mvis}
\end{figure}

The \textsc{DeepTau} algorithm has received major updates at the start of Run 3. This new version~\cite{CMS:2025kgf}, labeled v2.5, was optimized in terms of dataset preparation and hyperparameter optimization with respect to its predecessor (v2.1). The major update is a domain adaptation subnetwork, tasked with ensuring that the \textsc{DeepTau} performance is unaffected by potential mismodelling in the Monte Carlo simulation used for its training. This results in an improved simulation-to-data agreement for the \textsc{DeepTau} classifiers, as exemplified in Figs.~\ref{fig:domainadapt} and ~\ref{fig:2024-5mvis}.


Fig.~\ref{fig:domainadapt} compares the against-jet classifier of \textsc{DeepTau} with the domain adaptation being included or excluded, while Fig.~\ref{fig:2024-5mvis} shows the resulting simulation-to-data agreement during 2024 and in the first half of 2025~\cite{CMS-DP-2025-074}, presenting the invariant mass of a muon and \tauh system in the absence of dedicated calibrations on the \tauh candidate, underlying the accuracy of the modelling of the \textsc{DeepTau}v2.5 algorithm.

\vspace{-0.3cm}
\section{Alternative Strategies for Tau Identification}
\label{sec:3taggers}

\begin{figure}[h]
    \centering
    \includegraphics[width=0.35\linewidth]{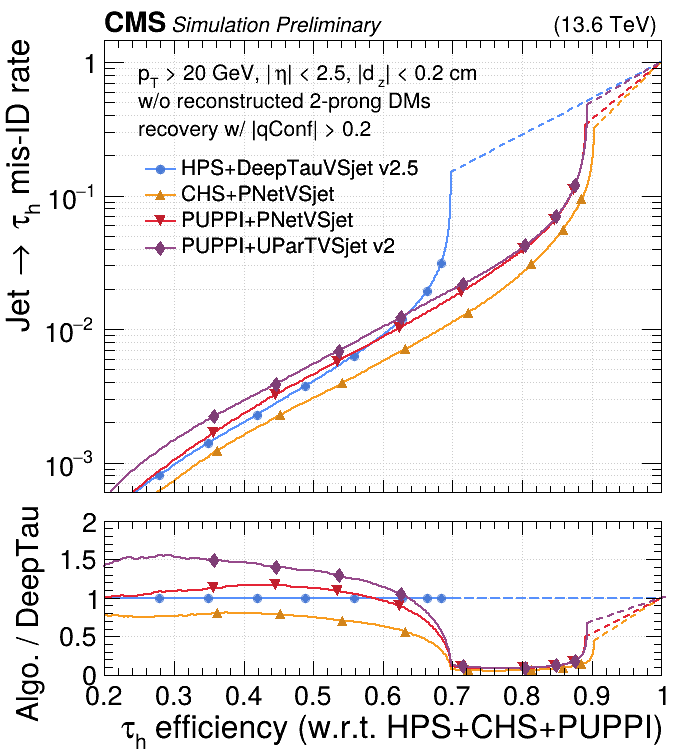}
    \includegraphics[width=0.35\linewidth]{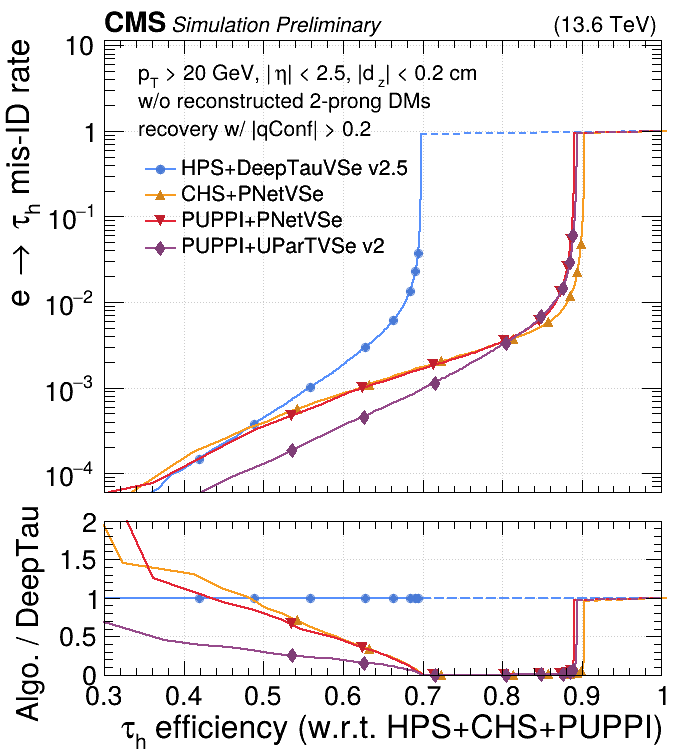}
    \caption{ROC curves illustrating the performance of the \textsc{DeepTau}, \textsc{PNet}, and \textsc{UParT} algorithms to discriminate quark and gluon jets (left) and electrons (right) misidentified as \tauh candidates~\cite{CMS-DP-2025-073}. The efficiencies are evaluated taking into account the preceding reconstruction and pileup mitigation algorithms used, labeled HPS, CHS, and PUPPI~\cite{CMS:2018jrd,CMS:2020ebo,Bertolini:2014bba,CMS-DP-2024-043}.}
    \label{fig:comparison}
\end{figure}

An alternative strategy for the identification of genuine \tauh candidates developed in CMS: the simultaneous classification of jet origin, treating the \tauh candidates as jets. This ``unified jet tagging'' philosophy provides a few advantages, such as an inherent modelling of the correlation between the identification of different objects, an easier optimization for the selection efficiency to prioritize the rejection of specific sources of contamination, and a general concentration of computing resources. Two algorithms have been introduced: \textsc{PNet}~\cite{PhysRevD.101.056019}, a graph neural network representing jets as a “particle cloud”, and \textsc{UParT}~\cite{Qu:2022mxj}, which relies on a transformer architecture.
As the jet classification performed by these algorithms is not limited to \tauh candidates selected by HPS, a dedicated comparison~\cite{CMS-DP-2025-073} that accounts for different reconstruction and pileup suppression techniques~\cite{CMS:2020ebo,Bertolini:2014bba,CMS-DP-2024-043}, was needed to estimate their efficiency and background rejection accurately. Fig.~\ref{fig:comparison} shows the rejection of jets and electrons of the presented algorithms as a function of the identification for genuine \tauh candidates. The algorithms perform differently across misidentified objects, highlighting how exploring different machine learning architectures can help improve the identification of \tauh candidates.

\vspace{-0.3cm}
\section{Taus in Nonstandard Topologies}
\label{sec:4nonstandard}

Additional algorithms have been deployed in CMS to target topologies not adequately covered by the standard reconstruction and identification techniques:
\begin{tight_itemize}
    \item \textsc{DisTau}~\cite{CMS-DP-2024-053}, a \textsc{PNet}-like algorithm targeting \tauh candidates produced displaced from the primary interaction vertex;
    \item \textsc{Boosted DeepTau}~\cite{CMS-DP-2025-047}, a CNN algorithm using the same architecture as \textsc{DeepTau} dedicated to \tauh candidates produced as part of a boosted $\tau\tau$ system where the decay products of the two tau leptons overlap;
    \item a modified version of HPS dedicated to low-\pt \tauh candidates recorded in the scouting data stream~\cite{CMS:2024zhe}, accompanied by a dedicated energy flow neural network named \textsc{TauNet}~\cite{Collaboration:2905110}.
\end{tight_itemize}

\begin{figure}[h]
    \centering
    \includegraphics[width=0.27\linewidth]{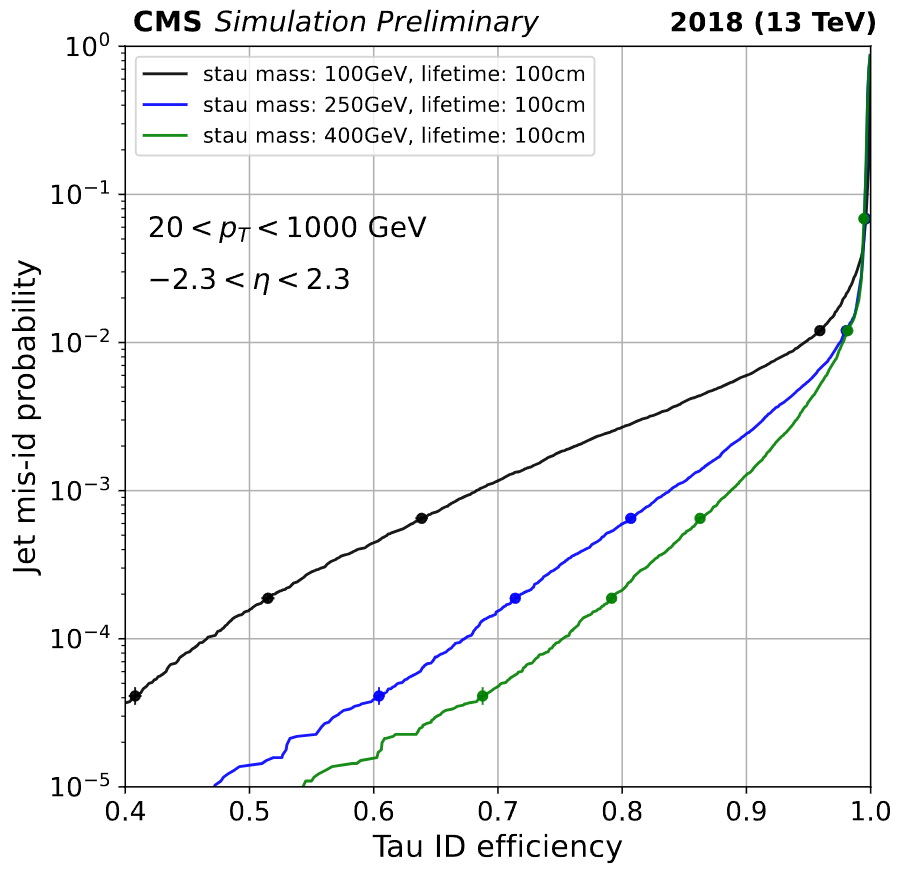}
    \includegraphics[width=0.27\linewidth]{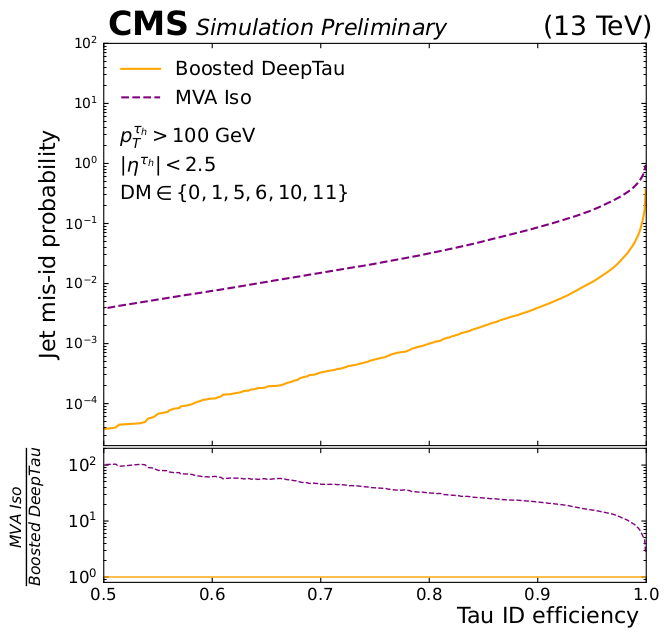}
    \includegraphics[width=0.33\linewidth]{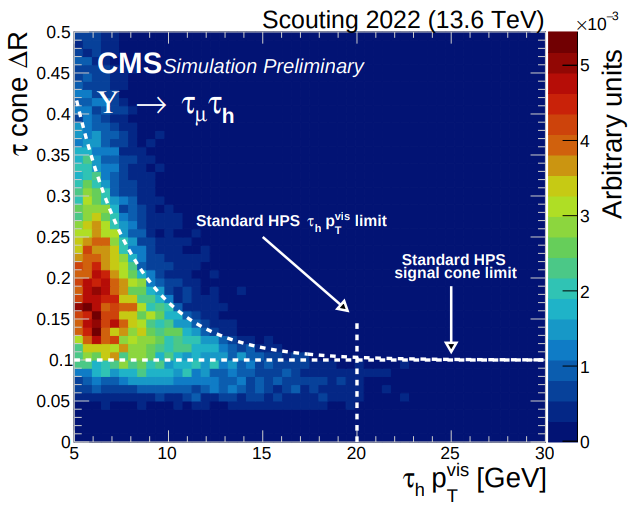}
    \caption{Left and middle: Jet rejection vs \tauh identification efficiency for the \textsc{DisTau} graph network at high displacement~\cite{CMS-DP-2024-053}, and for the \textsc{Boosted DeepTau} CNN algorithm at high \ptns~\cite{CMS-DP-2025-047}, respectively. Right: calibration of the low-\pt \tauh reconstruction algorithm using the $\Upsilon\rightarrow\tau_\mu$\tauh resonance~\cite{Collaboration:2905110}.}
    \label{fig:nonstd}
\end{figure}

Representative distributions to showcase these dedicated algorithms are presented in Fig.~\ref{fig:nonstd}

\vspace{-0.3cm}
\section{Tau Identification at the HLT}
\begin{figure}[h]
    \centering
    \includegraphics[width=0.335\linewidth]{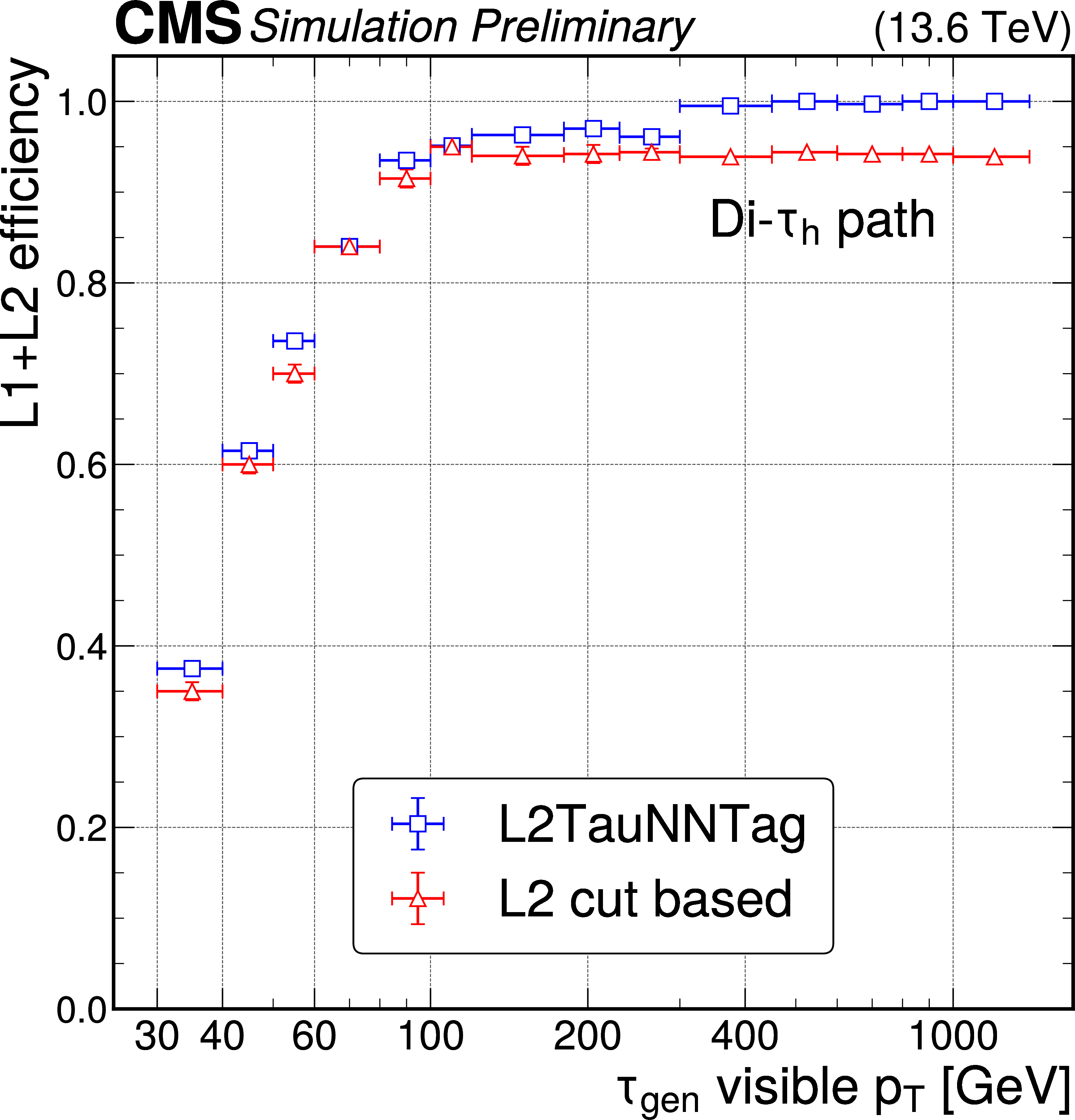}
    \includegraphics[width=0.35\linewidth]{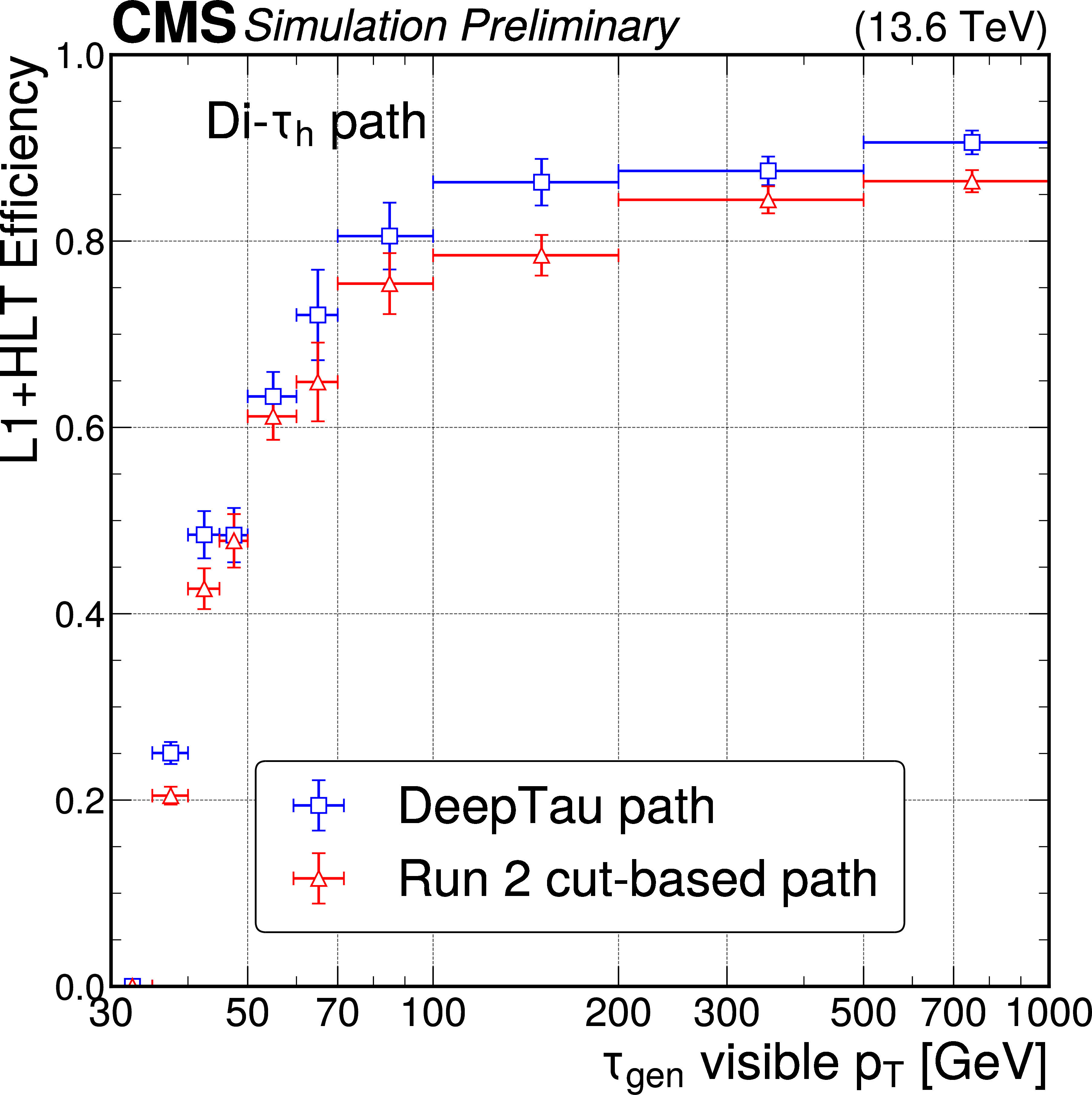}
    \caption{Efficiency as a function of the \tauh candidate \pt evaluated on the di-\tauh trigger algorithm at the L1+L2 stage (left) and with the full L1+HLT system (right)~\cite{CMS-PAS-TAU-24-002}.}
    \label{fig:trigger}
\end{figure}

With the exception of the scouting data flow, which relies only on information available at the hardware level trigger~\cite{CMS:2020cmk} of the CMS apparatus, proton-proton collisions are recorded based on a software-based high level trigger (HLT) system~\cite{CMS:2016ngn,CMS:2024aqx}. Run 3 saw the deployment of improved neural-network-based algorithms to identify \tauh candidates at different stages of the HLT reconstruction~\cite{CMS-PAS-TAU-24-002}: \textsc{L2TauNNTag} and an online version of \textsc{DeepTau}. Their relative improvement with respect to the cut-based algorithms used in Run 2 is shown in Fig.~\ref{fig:trigger} for the trigger algorithm that requires the simultaneous identification of two \tauh candidates.

\vspace{-0.3cm}

\section{Summary}

These proceedings provide an overview of recent advancements in reconstruction and identification techniques for hadronically decaying tau leptons deployed by the CMS Collaboration. These advancements both increase the overall efficiency in identifying tau leptons and widen the scope of physics analyses to previously unaddressed topologies.

\newpage

\bibliographystyle{cms_unsrt}
\bibliography{references}

@article{CMS:2008xjf,
    author = {{CMS} Collaboration},
    collaboration = "CMS",
    title = "{The CMS Experiment at the CERN LHC}",
    doi = "10.1088/1748-0221/3/08/S08004",
    journal = "JINST",
    volume = "3",
    pages = "S08004",
    year = "2008"
}

@article{HIG-20-006,
    author = {{CMS} Collaboration},
    collaboration = "CMS",
    title = "{Analysis of the $CP$ structure of the Yukawa coupling between the Higgs boson and $\tau$ leptons in proton-proton collisions at $ \sqrt{s} $ = 13 TeV}",
    eprint = "2110.04836",
    archivePrefix = "arXiv",
    primaryClass = "hep-ex",
    reportNumber = "CMS-HIG-20-006, CERN-EP-2021-189",
    doi = "10.1007/JHEP06(2022)012",
    journal = "JHEP",
    volume = "06",
    pages = "012",
    year = "2022"
}

@article{Cardini:2022qqy,
    author = "Cardini, Andrea",
    title = "{Methodologies to Measure the $\mathcal {CP}$ Structure of the Higgs Yukawa Coupling to Tau Leptons}",
    doi = "10.3390/universe8050256",
    journal = "Universe",
    volume = "8",
    number = "5",
    pages = "256",
    year = "2022"
}

@article{SMP-18-010,
    author = {{CMS} Collaboration},
    collaboration = "CMS",
    title = "{Measurement of the {\ensuremath{\tau}} lepton polarization in Z boson decays in proton-proton collisions at $ \sqrt{s} $ = 13 TeV}",
    eprint = "2309.12408",
    archivePrefix = "arXiv",
    primaryClass = "hep-ex",
    reportNumber = "CMS-SMP-18-010, CERN-EP-2023-169",
    doi = "10.1007/JHEP01(2024)101",
    journal = "JHEP",
    volume = "01",
    pages = "101",
    year = "2024"
}

@article{HIG-21-001,
    author = {{CMS} Collaboration},
    collaboration = "CMS",
    title = "{Searches for additional Higgs bosons and for vector leptoquarks in $\tau\tau$ final states in proton-proton collisions at $\sqrt{s}$ = 13 TeV}",
    eprint = "2208.02717",
    archivePrefix = "arXiv",
    primaryClass = "hep-ex",
    reportNumber = "CMS-HIG-21-001, CERN-EP-2022-137",
    doi = "10.1007/JHEP07(2023)073",
    journal = "JHEP",
    volume = "07",
    pages = "073",
    year = "2023"
}

@article{HIG-16-043,
    author = {{CMS} Collaboration},
    collaboration = "CMS",
    title = "{Observation of the Higgs boson decay to a pair of $\tau$ leptons with the CMS detector}",
    eprint = "1708.00373",
    archivePrefix = "arXiv",
    primaryClass = "hep-ex",
    reportNumber = "CMS-HIG-16-043, CERN-EP-2017-181",
    doi = "10.1016/j.physletb.2018.02.004",
    journal = "Phys. Lett. B",
    volume = "779",
    pages = "283",
    year = "2018"
}

@article{HIG-19-010,
    author = {{CMS} Collaboration},
    collaboration = "CMS",
    title = "{Measurements of Higgs boson production in the decay channel with a pair of $\tau $ leptons in proton{\textendash}proton collisions at $\sqrt{s}=13$ TeV}",
    eprint = "2204.12957",
    archivePrefix = "arXiv",
    primaryClass = "hep-ex",
    reportNumber = "CMS-HIG-19-010, CERN-EP-2022-027",
    doi = "10.1140/epjc/s10052-023-11452-8",
    journal = "Eur. Phys. J. C",
    volume = "83",
    number = "7",
    pages = "562",
    year = "2023"
}

@techreport{PAS-EXO-24-012,
      author = {{CMS} Collaboration},
      collaboration = "CMS",
      title         = "{Search for a low mass resonance decaying to $\tau\tau$
                       using data collected with a dedicated high-rate data
                       stream}",
      reportNumber  = "CMS-PAS-EXO-24-012",
      number  = "CMS-PAS-EXO-24-012",
      year          = "2025",
	   type          = {{CMS Physics Analysis Summary}},
      url           = "https://cds.cern.ch/record/2931236",
}

@techreport{PAS-EXO-24-020,
      author = {{CMS} Collaboration},
      collaboration = "CMS",
      title         = "{Search for the pair production of long-lived
                       supersymmetric partners of the tau lepton in proton-proton
                       collisions at sqrt{s}=13 TeV}",
      reportNumber  = "CMS-PAS-EXO-24-020",
	   type          = {{CMS Physics Analysis Summary}},
	   number        = "CMS-PAS-EXO-24-020",
      year          = "2025",
      url           = "https://cds.cern.ch/record/2938553",
}

@article{HIN-21-009,
    author = {{CMS} Collaboration},
    collaboration = "CMS",
    title = "{Observation of $\tau$ lepton pair production in ultraperipheral lead-lead collisions at $\sqrt{s_\mathrm{NN}}$ = 5.02 TeV}",
    eprint = "2206.05192",
    archivePrefix = "arXiv",
    primaryClass = "nucl-ex",
    reportNumber = "CMS-HIN-21-009, CERN-EP-2022-098",
    doi = "10.1103/PhysRevLett.131.151803",
    journal = "Phys. Rev. Lett.",
    volume = "131",
    pages = "151803",
    year = "2023"
}

@techreport{CMS-PAS-HIN-24-011,
    author = {{CMS} Collaboration},
      collaboration = "CMS",
      title         = "{Measurement of the tau g-2 factor in ultraperipheral PbPb
                       collisions recorded by the CMS experiment}",
      institution   = "CERN",
      reportNumber  = "CMS-PAS-HIN-24-011",
	   type          = {{CMS Physics Analysis Summary}},
      number  = "CMS-PAS-HIN-24-011",
      year          = "2024",
      url           = "https://cds.cern.ch/record/2912969",
}

@article{CMS:2023gfb,
    author = {{CMS} Collaboration},
    collaboration = "CMS",
    title = "Development of the {CMS} detector for the {CERN LHC Run 3}",
    eprint = "2309.05466",
    archivePrefix = "arXiv",
    primaryClass = "physics.ins-det",
    reportNumber = "CMS-PRF-21-001, CERN-EP-2023-136",
    doi = "10.1088/1748-0221/19/05/P05064",
    journal = "JINST",
    volume = "19",
    pages = "P05064",
    year = "2024"
}

@article{CMS:2017yfk,
    author = {{CMS} Collaboration},
      title          = "Particle-flow reconstruction and global event description with the {CMS} detector",
      collaboration  = "CMS",
      journal     = "JINST",
      volume    = "12",
      year         = "2017",
      pages      = "P10003",
      doi           = "10.1088/1748-0221/12/10/P10003",
      eprint         = "1706.04965",
      archivePrefix  = "arXiv",
      primaryClass   = "physics.ins-det",
      reportNumber   = "CMS-PRF-14-001, CERN-EP-2017-110",
      SLACcitation   = "%%CITATION = ARXIV:1706.04965;%%",
}

@article{CMS:2020uim,
    author = {{CMS} Collaboration},
    collaboration = "CMS",
    title         = "Electron and photon reconstruction and identification with the {CMS} experiment at the {CERN} {LHC}",
    eprint        = "2012.06888",
    journal       = "JINST",
    volume        = "16",
    pages         = "P05014",
    year          = "2021",
    archivePrefix = "arXiv",
    primaryClass  = "hep-ex",
    reportNumber  = "CMS-EGM-17-001, CERN-EP-2020-219",
    doi           = "10.1088/1748-0221/16/05/P05014"
}

@article{CMS:2018rym,
    author = {{CMS} Collaboration},
    collaboration = "CMS",
    title = "Performance of the {CMS} muon detector and muon reconstruction with proton-proton collisions at $\sqrt{s}=13$\,{TeV}",
    eprint = "1804.04528",
    archivePrefix = "arXiv",
    primaryClass = "physics.ins-det",
    reportNumber = "CMS-MUO-16-001, CERN-EP-2018-058",
    doi = "10.1088/1748-0221/13/06/P06015",
    journal = "JINST",
    volume = "13",
    pages = "P06015",
    year = "2018"
}

@article{CMS:2014pgm,
    author = {{CMS} Collaboration},
    collaboration = "CMS",
    title = "Description and performance of track and primary-vertex reconstruction with the {CMS} tracker",
    eprint = "1405.6569",
    archivePrefix = "arXiv",
    primaryClass = "physics.ins-det",
    reportNumber = "CMS-TRK-11-001, CERN-PH-EP-2014-070",
    doi = "10.1088/1748-0221/9/10/P10009",
    journal = "JINST",
    volume = "9",
    pages = "P10009",
    year = "2014"
}

@article{CMS:2018jrd,
    author = {{CMS} Collaboration},
    collaboration = "CMS",
    title = "Performance of reconstruction and identification of $\tau$ leptons decaying to hadrons and $\nu_\tau$ in pp collisions at $\sqrt{s}=13$\,{TeV}",
    eprint = "1809.02816",
    archivePrefix = "arXiv",
    primaryClass = "hep-ex",
    reportNumber = "CMS-TAU-16-003, CERN-EP-2018-229",
    doi = "10.1088/1748-0221/13/10/P10005",
    journal = "JINST",
    volume = "13",
    pages = "P10005",
    year = "2018"
}

@article{CMS:2022prd,
    author = {{CMS} Collaboration},
    collaboration = "CMS",
    title = "Identification of hadronic tau lepton decays using a deep neural network",
    eprint = "2201.08458",
    archivePrefix = "arXiv",
    primaryClass = "hep-ex",
    reportNumber = "CMS-TAU-20-001, CERN-EP-2021-257",
    doi = "10.1088/1748-0221/17/07/P07023",
    journal = "JINST",
    volume = "17",
    pages = "P07023",
    year = "2022"
}

@article{CMS:2025kgf,
    author = {{CMS} Collaboration},
    collaboration = "CMS",
    title = "{Identification of tau leptons using a convolutional neural network with domain adaptation}",
    eprint = "2511.05468",
    archivePrefix = "arXiv",
    primaryClass = "hep-ex",
    reportNumber = "CMS-TAU-24-001, CERN-EP-2025-233",
    year = "2025",
    note = "Submitted to JINST"
}

@techreport{CMS-DP-2025-074,
    author = {{CMS} Collaboration},
      collaboration = "CMS",
      title         = "{Validation of tau identification algorithms in Run 3 data
                       and simulation}",
      year          = "2025",
      url           = "https://cds.cern.ch/record/2946446",
	type        = "{CMS} Detector Performance Note",
	number     = "CMS-DP-2025-074"
}

@techreport{CMS-DP-2025-073,
    author = {{CMS} Collaboration},
      collaboration = "CMS",
      title         = "{Comparison of the performance of tau reconstruction and
                       identification algorithms in Run 3}",
      year          = "2025",
      url           = "https://cds.cern.ch/record/2946445",
	type        = "{CMS} Detector Performance Note",
	number     = "CMS-DP-2025-073"
}

@article{PhysRevD.101.056019,
  title = {Jet tagging via particle clouds},
  author = {Qu, Huilin and Gouskos, Loukas},
  journal = {Phys. Rev. D},
  volume = {101},
  issue = {5},
  pages = {056019},
  numpages = {11},
  year = {2020},
  publisher = {American Physical Society},
  doi = {10.1103/PhysRevD.101.056019},
  url = {https://link.aps.org/doi/10.1103/PhysRevD.101.056019}
}

@article{Qu:2022mxj,
    author = "Qu, Huilin and Li, Congqiao and Qian, Sitian",
    title = "{Particle Transformer for Jet Tagging}",
    eprint = "2202.03772",
    archivePrefix = "arXiv",
    primaryClass = "hep-ph",
    year = "2022"
}

@techreport{CMS-DP-2024-053,
    AUTHOR      = "{CMS Collaboration}",
    collaboration = "CMS",
    title         = "{Tau lepton identification in displaced topologies using machine learning at CMS}",
    year          = "2024",
    url           = "https://cds.cern.ch/record/2904366",
    TYPE        = "CMS Detector Performance Summary",
    NUMBER      = "CMS-DP-2024-053",
}

@techreport{CMS-DP-2025-047,
    AUTHOR      = "{CMS Collaboration}",
      collaboration = "CMS",
      title         = "{Performance of boosted tau lepton identification with
                       DeepTau Framework (\textsc{Boosted DeepTau})}",
      year          = "2025",
      url           = "https://cds.cern.ch/record/2941434",
    TYPE        = "CMS Detector Performance Summary",
    NUMBER      = "CMS-DP-2025-047",
}

@techreport{Collaboration:2905110,
    AUTHOR      = "{CMS Collaboration}",
      collaboration = "CMS",
      title         = "{Low transverse-momentum hadronic tau lepton
                       reconstruction performance in the Run 3 Scouting dataset}",
      institution   = "CERN",
      reportNumber  = "CMS-NOTE-2024-006, CERN-CMS-NOTE-2024-006",
      year          = "2024",
      url           = "https://cds.cern.ch/record/2905110",
    type = {{CMS Note}},
      number  = "CMS-NOTE-2024-006",
}

@article{CMS:2024zhe,
    author      = "{CMS Collaboration}",
    collaboration = "CMS",
    title = "{Enriching the physics program of the CMS experiment via data scouting and data parking}",
    eprint = "2403.16134",
    archivePrefix = "arXiv",
    primaryClass = "hep-ex",
    reportNumber = "CMS-EXO-23-007, CERN-EP-2024-068",
    doi = "10.1016/j.physrep.2024.09.006",
    journal = "Phys. Rept.",
    volume = "1115",
    pages = "678",
    year = "2025"
}

@article{CMS:2020cmk,
    author      = "{CMS Collaboration}",
    collaboration = "CMS",
    title = "Performance of the {CMS} {Level-1} trigger in proton-proton collisions at $\sqrt{s} = 13$\,{TeV}",
    journal = "JINST",
    volume = "15",
    pages = "P10017",
    year = "2020",
    doi = "10.1088/1748-0221/15/10/P10017",
    eprint = "2006.10165",
    archivePrefix = "arXiv",
    primaryClass = "hep-ex",
    reportNumber = "CMS-TRG-17-001, CERN-EP-2020-065",
}

\end{document}